\documentclass[aps,prl,twocolumn,superscriptaddress,nofootinbib]{revtex4-2}

\usepackage{amsmath,amssymb,bm,graphicx}
\usepackage[dvipsnames]{xcolor}
\usepackage{hyperref}
\usepackage{ragged2e}
\hypersetup{
colorlinks=true,
urlcolor=blue,
citecolor=blue,
linkcolor=blue
}

\newcommand{\rd}{r_{\mathrm d}}
\newcommand{\rdf}{r^{\mathrm{fid}}_{\mathrm d}}
\newcommand{\dm}{D_{\mathrm M}}
\newcommand{\dmf}{D^{\mathrm{fid}}_{\mathrm M}}
\renewcommand{\dh}{D_{\mathrm H}}
\newcommand{\dhf}{D^{\mathrm{fid}}_{\mathrm H}}
\newcommand{\apar}{\alpha_\parallel}
\newcommand{\aperp}{\alpha_\perp}

\newcommand{\nn}{\nonumber}

\begin{document}

\title{A calibration-free null test from anisotropic BAO}

\author{Domenico Sapone}
\email{domenico.sapone@uchile.cl}
\affiliation{Departamento de Física, FCFM, Universidad de Chile, Santiago, Chile}

\author{Savvas Nesseris}
\email{savvas.nesseris@csic.es}
\affiliation{Instituto de Física Teórica UAM-CSIC, Universidad Autónoma de Madrid, Cantoblanco, 28049 Madrid, Spain}

\date{\today}

\begin{abstract}
Baryon acoustic oscillation (BAO) analyses usually report the anisotropic shift parameters $\aperp(z)$ and $\apar(z)$ relative to a fiducial cosmology, and these quantities are primarily used for cosmological parameter inference. Here we show that they can also be used to construct a direct internal consistency test of the background geometry. In particular, we derive a new null test of flat Friedmann-Lema\^itre-Robertson-Walker (FLRW) geometry written entirely in terms of the reported BAO shift parameters. The test is calibration free: the sound-horizon ratio $\rd/\rdf$ cancels identically, so the relation is independent of the absolute BAO scale. We also derive a calibration-free reconstruction of the deceleration parameter $q(z)$ from the radial BAO sector. Applying these results to anisotropic DESI DR2 BAO measurements, we find no evidence for a breakdown of the flat-FLRW distance relation within current uncertainties. Our results show that anisotropic BAO measurements already provide a nontrivial internal geometric consistency test before performing any model fit.
\end{abstract}

\maketitle

\paragraph{Introduction.}
Baryon acoustic oscillation (BAO) measurements have become one of the central probes of late-time cosmology, providing robust constraints on the expansion history through characteristic distance scales measured in galaxy clustering \cite{Weinberg:2013agg,SDSS:2005xqv,2dFGRS:2005yhx,Blake:2011en,BOSS:2012dmf,eBOSS:2015jyv,DESI:2025zgx}. In anisotropic BAO analyses, the data are commonly summarized in terms of the transverse and radial shift parameters, $\aperp(z)$ and $\apar(z)$, defined with respect to a chosen fiducial cosmology. These are then used as inputs for model comparison and parameter inference.

Here we take a different point of view. Rather than asking which cosmological model best fits the BAO distances, we ask a more basic question: do the BAO-derived transverse and radial distances behave as distances in a flat-FLRW spacetime at all? This distinction is important. A global parameter fit can absorb mild inconsistencies into shifted cosmological parameters, whereas a null test directly probes whether the measured quantities satisfy a relation that must hold if the underlying geometric framework is correct.

Our main result is a new null test of flat-FLRW geometry constructed directly from the reported anisotropic BAO shift parameters. The test is calibration free: any direct information on the sound-horizon at the baryon drag redshift cancels identically, so a failure of the null relation cannot be attributed simply to the choice of the BAO scale.

We also derive a calibration-free reconstruction of the deceleration parameter $q(z)$ from the radial BAO sector. Furthermore, since combined DESI DR2 \cite{DESI:2025zgx,DES:2025sig,Ong:2026tta} analyses have been interpreted as favoring evolving dark energy at roughly the $3\sigma$ level in parametrized fits such as CPL \cite{Chevallier:2000qy, Linder:2002et}, we also explore whether the BAO sector alone can support an effective reconstruction $w(z)$.

These results complement the broader literature on cosmological consistency tests and null diagnostics \cite{Clarkson:2007pz,Clarkson:2012bg,Nesseris:2010ep,Heavens:2011mr,Laurent:2016eqo,Chiang:2017yrq,Nesseris:2014mfa,Nesseris:2014qca,Andrade:2021njl, Seikel:2012cs, vonMarttens:2018iav, vonMarttens:2020apn, Garcia-Bellido:2008xmz,Millard:2026wnd,Sapone:2014nna}. The novelty here is that the construction is written directly in terms of the anisotropic BAO shift parameters reported by the survey and our two main results are calibration free. 

Applying these relations to current anisotropic BAO data from DESI DR2 \cite{DESI:2025zgx}, we find that the data are consistent with the flat-FLRW distance relation within present uncertainties and $w(z)$ does not show a significant deviation from $w=-1$.

\paragraph{BAO shifts and the flat-FLRW null relation.}
The anisotropic BAO shift parameters are defined as
\begin{equation}
\aperp(z)=\frac{\dm(z)/\rd}{\dmf(z)/\rdf},
\qquad
\apar(z)=\frac{\dh(z)/\rd}{\dhf(z)/\rdf},
\label{eq:alphadef}
\end{equation}
where $\dm(z)=(1+z)D_A(z)$ is the transverse comoving distance, $\dh(z)=c/H(z)$ is the Hubble distance. Here $\rd$ is the comoving sound horizon at the baryon-drag epoch, which sets the physical BAO standard ruler. In galaxy clustering analysis, the anisotropic correlation is measured in fiducial coordinates and compared with a rescaled template, schematically the 2-point correlation function is $\xi(r_\parallel,r_\perp)=\xi_\mathrm{fid}(\apar\,r_\parallel,\aperp\,r_\perp)$, so that $\aperp$ and $\apar$  quantify the radial and transverse shift of the BAO feature relative to the fiducial cosmology.

In a spatially flat-FLRW spacetime, the transverse comoving distance equals the line-of-sight comoving distance,
\begin{equation}
D_\mathrm{C}(z)=D_M(z)=c\int_0^z \frac{d\tilde z}{H(\tilde z)},
\end{equation}
and therefore obeys the exact identity
\begin{equation}
D_M'(z)=D_H(z),
\label{eq:DMDH}
\end{equation}
where a prime denotes derivative with respect to redshift.

Using Eq.~\eqref{eq:alphadef}, we may write
\begin{eqnarray}
D_M(z)&=&\frac{\rd}{\rdf}\,\dmf(z)\,\aperp(z)\,,\label{eq:DMalpha}
\\
D_H(z)&=&\frac{\rd}{\rdf}\,\dhf(z)\,\apar(z)\,.
\label{eq:DHalpha}
\end{eqnarray}
Substituting these expressions into Eq.~(\ref{eq:DMDH}), the prefactor $\rd/\rdf$ cancels identically:
\begin{equation}
\left[\dmf(z)\,\aperp(z)\right]'=\dhf(z)\,\apar(z).
\label{eq:master}
\end{equation}
For a flat fiducial cosmology, one also has $(\dmf)'=\dhf$, and Eq.~(\ref{eq:master}) reduces to
\begin{equation}
\mathcal{C}(z)\equiv
\dmf(z)\,\frac{\aperp'(z)}{\aperp(z)}
-\dhf(z)\left[\frac{\apar(z)}{\aperp(z)}-1\right]=0.
\label{eq:Cz}
\end{equation}
Equation~(\ref{eq:Cz}) is the main result of this work. It is an exact null relation for any flat-FLRW cosmology, independent of the detailed dark-energy model, and independent of the absolute BAO calibration scale.

The condition $\mathcal{C}(z)=0$ does not require $\aperp=\apar=1$; percent-level shifts of the BAO shift parameters away from unity merely indicate that the preferred distances differ from those of the fiducial cosmology. Conversely, a statistically significant departure from $\mathcal{C}(z)=0$ would signal an inconsistency between the transverse and radial BAO sectors under the flat-FLRW distance relation. Such a failure could arise from spatial curvature, a breakdown of the assumed background-distance relation~\cite{Euclid:2020ojp}, residual observational systematics, or inconsistencies in the extraction of the anisotropic BAO distances~\cite{Chen:2024tfp}.

A numerically more stable integrated form follows directly from Eq.~(\ref{eq:master}):
\begin{equation}
\mathcal{C}_{\rm int}(z)\equiv
\aperp(z)
-
\frac{
\dmf(z_0)\,\aperp(z_0)+\int_{z_0}^{z}d\tilde z\,\dhf(\tilde z)\,\apar(\tilde z)
}{
\dmf(z)
},
\label{eq:Cint}
\end{equation}
which also vanishes identically in flat-FLRW and avoids numerical derivatives of the transverse BAO sector.

\paragraph{Fiducial-cosmology remapping.}
The BAO shift parameters are not themselves direct observables; they are inferred relative to a fiducial cosmology used to convert observed redshifts and angles into distances \cite{Padmanabhan:2008ag,Xu:2012fw,BOSS:2013rlg,BOSS:2016wmc}. Nevertheless, the physical BAO information resides in the combinations $\dm/\rd$ and $\dh/\rd$, which are independent of the chosen fiducial model. As a result, the reported shift parameters can be remapped exactly from one fiducial cosmology $A$ to another one $B$:
\begin{align}
\apar(B) &= \apar(A)\,
\frac{\dhf(A)}{\dhf(B)}
\frac{\rdf(B)}{\rdf(A)},
\\
\aperp(B) &= \aperp(A)\,
\frac{\dmf(A)}{\dmf(B)}
\frac{\rdf(B)}{\rdf(A)}.
\label{eq:remap}
\end{align}
This mapping is exact and does not assume any specific true cosmology \cite{Carter:2019ulk}. If both fiducial cosmologies are flat, the null condition $\mathcal{C}(z)=0$ is preserved under the transformation, showing that the test is not an artifact of one particular fiducial choice.

\paragraph{A calibration-free reconstruction of $q(z)$.}
The same framework also gives access to a simple kinematic quantity. The deceleration parameter is defined by
\begin{equation}
q(z)\equiv -1+(1+z)\frac{H'(z)}{H(z)}
       = -1-\frac{d\ln D_H}{d\ln(1+z)}.
\label{eq:qdef}
\end{equation}
Using Eq.~(\ref{eq:DHalpha}) for $D_H(z)$, the factor $\rd/\rdf$ cancels once again, yielding
\begin{eqnarray}
q(z)&=&
-1-(1+z)\left[
\frac{(\dhf)'(z)}{\dhf(z)}
+
\frac{\apar'(z)}{\apar(z)}
\right]\,,
\label{eq:qalpha}\\
q(z) &=& q_{\rm fid}(z) - (1+z)\,\frac{\alpha_\parallel'}{\alpha_\parallel}\,.
\end{eqnarray}
where $q_{\rm fid} \equiv -1 - (1+z)(\dhf)'(z)/\dhf(z)$.
Thus, $q(z)$ can be reconstructed directly from the redshift dependence of the radial BAO shift parameter without any dependence on the absolute sound horizon. Unlike $\mathcal{C}(z)$, this is not a null test: its value depends on the expansion history. Still, it provides a calibration-free kinematic diagnostic derived entirely from the radial BAO sector.

\begin{figure}[t]
\includegraphics[width=\columnwidth]{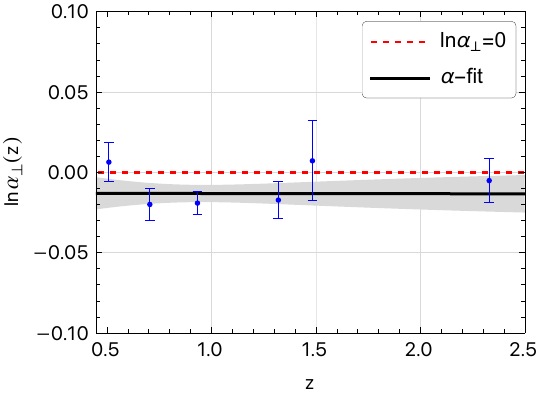}
\includegraphics[width=\columnwidth]{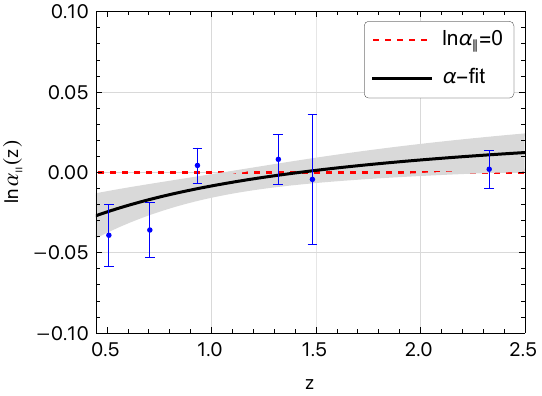}
\caption{Anisotropic BAO shift parameters from DESI DR2. Blue points denote the binned measurements of $\aperp(z)$ (top) and $\apar(z)$ (bottom). The black curves show a smooth two-parameter fit to $\ln\aperp$ and $\ln\apar$, with the shaded region indicating the propagated $68.3\%$ uncertainty.}
\label{fig:alpha}
\end{figure}
\paragraph{A diagnostic reconstruction of $w(z)$.}
The same framework can also be extended to infer an effective dark-energy equation of state $w(z)$ directly from the BAO sector. Unlike $\mathcal{C}(z)$ and $q(z)$, however, this reconstruction is not calibration free: it depends on an overall late-time normalization through $(h\,r_{\rm d})$ and on matter density $\Omega_\mathrm{m,0}$; it is  not independent of the absolute BAO scale and on extra input parameters. In this sense, $w(z)$ is not a null test, but a more model-dependent diagnostic of the expansion history. In a flat cosmology, it may be written as \cite{Huterer:1998qv}
\begin{align}
w(z)=
\frac{
-3-2(1+z)\left[
\frac{(\dhf)'}{\dhf}
+\frac{\alpha_\parallel'}{\alpha_\parallel}
\right]
}{
1-\frac{10^4}{c^2}
\left(\frac{h\,\rd}{\rdf}\right)^2
\left(\dhf\alpha_\parallel\right)^2
\Omega_\mathrm{m,0}\,A_{\rm eq}(z)\,(1+z)^3
}\,,\label{eq:wz}
\end{align}
where $A_{\rm eq}=1+(1+z)/(1+z_{\rm eq})$ accounts for radiation corrections at high redshift and we assume the DESI DR2 best-fit values for $h\,\rd$ and $\Omega_\mathrm{m,0}$.

\paragraph{Application to DESI DR2.}
We apply Eqs.~(\ref{eq:Cz}), \eqref{eq:Cint}, \eqref{eq:qalpha}, and (\ref{eq:wz}) to the anisotropic BAO measurements reported by DESI DR2 \cite{DESI:2025zgx}. In the main analysis we use the  binned values of $\ln\aperp(z)$ and $\ln\apar(z)$ together with their covariance matrix. 
This choice is convenient for two reasons. 
First, the consistency relations derived in this work naturally involve logarithmic derivatives. 
Second, if one moves from $(\alpha_\perp,\alpha_\parallel)$ to $(\alpha_{\rm iso},\alpha_{\rm AP})$, the latter being the values actually published by DESI, the transformation becomes linear in logarithmic space:
\begin{equation}
\ln \alpha_{\rm iso}=\frac{2\ln\alpha_\perp+\ln\alpha_\parallel}{3}\,, 
\qquad
\ln \alpha_{\rm AP}=\ln\alpha_\parallel-\ln\alpha_\perp\,.\nn
\end{equation}
Uncertainties are propagated using standard linear error propagation using the full covariance of the data. 

For derivative-based quantities we estimate first derivatives on the discrete redshift grid using standard forward, backward, and central finite differences, and propagate uncertainties by linear covariance propagation. To test the robustness of the derivative-based reconstruction against noise amplification, we also consider a simple parametric fit for the logarithms of the anisotropic BAO shift parameters,
\begin{align}
\ln \alpha_\perp(z) &= \ln \alpha_{\perp,0}
+ \ln \alpha_{\perp,a}\,\frac{z}{1+z},
\label{eq:a-fit1}\\
\ln \alpha_\parallel(z) &= \ln \alpha_{\parallel,0}
+ \ln \alpha_{\parallel,a}\,\frac{z}{1+z},
\label{eq:a-fit2}
\end{align}
which provides a smooth interpolation across the observed redshift range and may be viewed as a low-order expansion around \(a=1\). We refer to this parametrization as the ``\(\alpha\)-fit''.

The fit is performed directly to the BAO data, including the full data covariance matrix, and therefore involves only four free parameters:
\[
\{
\ln \alpha_{\perp,0},\,
\ln \alpha_{\perp,a},\,
\ln \alpha_{\parallel,0},\,
\ln \alpha_{\parallel,a}
\}.
\]
From the best-fit parameters and their data covariance, we reconstruct the diagnostics discussed above and derive the corresponding \(68.3\%\) confidence regions. The input data and the resulting \(\alpha\)-fit are shown in Fig.~\ref{fig:alpha}.

\begin{figure}[t]
\includegraphics[width=\columnwidth]{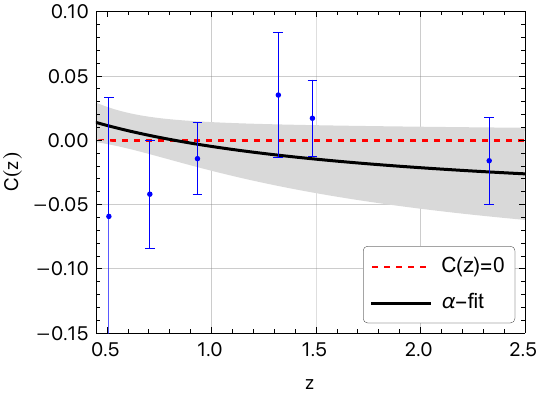}
\includegraphics[width=\columnwidth]{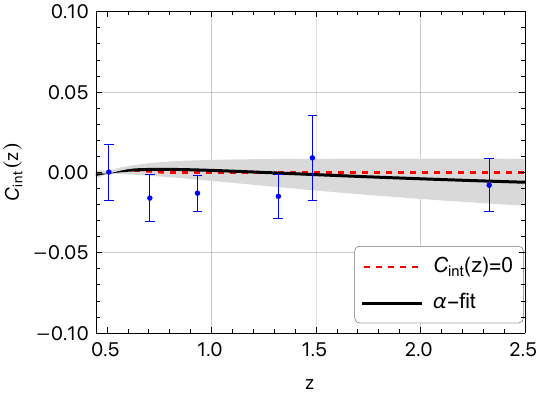}
\caption{Top: reconstruction of the flat-FLRW null test $\mathcal{C}(z)$ from anisotropic BAO data. Bottom: integrated version $\mathcal{C}_{\rm int}(z)$. In both cases the dashed red line indicates the null expectation.}
\label{fig:c-test}
\end{figure}

\paragraph{Results.-} The main empirical result, shown in Fig.~\ref{fig:c-test}, is that the reconstructed $\mathcal{C}(z)$ is consistent with zero across the DESI DR2 redshift range within current uncertainties. The same conclusion is reached using the integrated estimator $\mathcal{C}_{\rm int}(z)$, which is numerically more stable and less sensitive to finite-difference noise. We therefore find no evidence, at present precision, for an internal breakdown of the flat-FLRW distance relation in the anisotropic BAO sector.

\begin{figure}[t]
\includegraphics[width=\columnwidth]{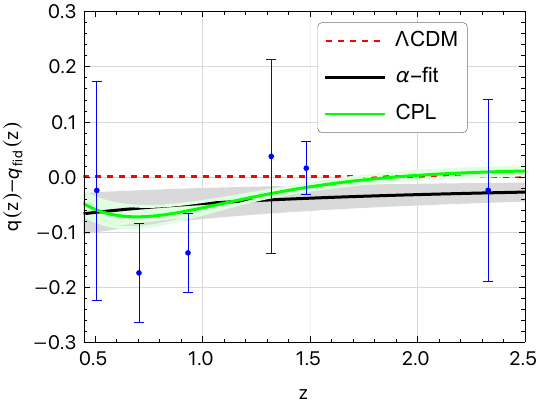}
\caption{Calibration-free reconstruction of the deceleration parameter $q(z)$ from the radial BAO sector. The dashed red line marks zero and the fiducial model; the green curve and black line show the DESI best fit and the $\alpha$-fit reconstruction respectively, with the shaded regions indicating the $68.3\%$ confidence interval.}
\label{fig:q-test}
\end{figure}

The reconstruction of the deceleration parameter is shown in Fig.~\ref{fig:q-test} in terms of the difference $q(z)-q_{\rm fid}(z)$, which directly highlights deviations from the fiducial kinematic evolution. Within the current uncertainties, the reconstructed points are consistent with zero across the full redshift range, indicating no significant departure from the fiducial late-time expansion history. This is consistent with the interpretation that present anisotropic BAO data do not yet require a deviation from the standard smooth background evolution at the level of this calibration-free kinematic diagnostic.

The DESI best-fit CPL prediction displays a stronger redshift dependence and follows the fluctuations of the reconstructed points more closely. However, given the size of the current error bars and the limited number of redshift bins, this behavior should not be over-interpreted as evidence for evolving dark energy. At present precision, the BAO-based reconstruction of $q(z)$ remains compatible both with the fiducial model and with smooth alternative expansion histories.

\begin{figure}[!t]
\centering
\includegraphics[width=0.95\linewidth]{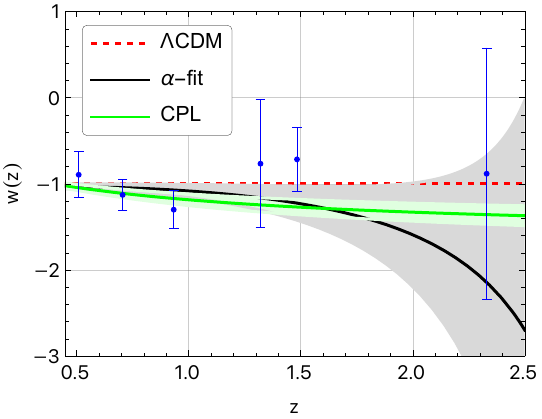}
\caption{\justifying BAO-based reconstruction of $w(z)$. The dashed red line marks $w=-1$; the green curve and black line show the DESI best fit and the $\alpha$-fit reconstruction respectively, with the shaded regions indicating the $68.3\%$ confidence interval.}
\label{fig:wz}
\end{figure}

Finally, as shown in Fig.~\ref{fig:wz}, the reconstructed $w(z)$ is consistent with $w=-1$ within the current uncertainties, but it exhibits substantially larger scatter than the corresponding reconstructions of $\mathcal{C}(z)$ and $q(z)$. This behavior is expected, since the inference of $w(z)$ involves derivatives of the radial BAO shift parameter as well as nonlinear combinations that amplify noise and covariance effects. At present precision, the BAO-based reconstruction of $w(z)$ should therefore be regarded mainly as an illustrative extension of the method rather than as robust evidence for dark-energy evolution.

\paragraph{Conclusions.—}
We have shown that anisotropic BAO shift parameters can be used for direct internal consistency tests of the background geometry. Our main result is the null relation in Eq.~\eqref{eq:Cz}, an exact calibration-free test of the flat-FLRW distance identity written directly in terms of the reported BAO shift parameters. Because the sound-horizon ratio cancels identically, this relation probes the internal consistency of the BAO geometric sector independently of the absolute BAO calibration scale.

Applied to current DESI DR2 anisotropic BAO measurements, the null test is consistent with zero within present uncertainties, and this conclusion is robust under exact remapping between flat fiducial cosmologies. We also derived a calibration-free reconstruction of the deceleration parameter from the radial BAO sector, showing that anisotropic BAO data already contain direct kinematic information before any global cosmological fit is performed.

The broader message is that anisotropic BAO measurements provide a nontrivial internal geometric consistency test of late-time cosmology. As future surveys improve the precision, redshift coverage, and control of anisotropic BAO covariances, the null relation introduced here can become a sharp test of the flat-FLRW framework itself.

Interestingly, both the calibration-free reconstruction of $q(z)-q_{\rm fid}(z)$ and the diagnostic reconstruction of $w(z)$ point to the same intermediate-redshift bins as carrying the largest leverage. In both cases, the second and third bins exhibit the largest departures from the fiducial expectation and appear to account for much of the apparent preference for a CPL-like evolution, which has been noted in the literature \cite{Sapone:2024ltl,Chudaykin:2024gol,Colgain:2024xqj}. While these departures are visible at the $1\,\sigma$ level, they are already consistent with $\Lambda$CDM once 
$2\,\sigma$ uncertainties are considered. At present precision, however, this indication remains suggestive rather than definitive.\\

\paragraph{Acknowledgements.—}
DS acknowledges financial support from Fondecyt Regular N. 1251339. 
SN acknowledges support from the research project PID2024-159420NB-C43, the Proyecto de Investigación SAFE25003 from the Consejo Superior de Investigaciones Científicas (CSIC), and the Spanish Research Agency (Agencia Estatal de Investigaci\'on) through the Grant IFT Centro de Excelencia Severo Ochoa No CEX2020-001007-S, funded by MCIN/AEI/10.13039/501100011033.

\bibliography{references}

\end{document}